\documentclass[twocolumn,prb,superscriptaddress,longbibliography]{revtex4-2}
\usepackage{graphicx}
\usepackage{color}
\usepackage{amsmath}
\usepackage{enumitem}
\usepackage{amssymb}
\usepackage{hyperref}
\usepackage[normalem]{ulem}
\usepackage{adjustbox}

\hypersetup{
	colorlinks = true,
	linkcolor = blue,
	citecolor = blue,
	urlcolor  = blue,
}

\newcommand{\be}{\begin{equation}}
	\newcommand{\ee}{\end{equation}}

\newcommand{\bea}{\begin{eqnarray}}
	\newcommand{\eea}{\end{eqnarray}}

\newcommand{\tr}{{\rm \, tr\,}}

\renewcommand{\vec}[1]{{\boldsymbol #1}}
\renewcommand{\epsilon}{\varepsilon}

\def\nn{\nonumber\\}

\begin{document}
	\title{RKKY quadratic and biquadratic spin-spin interactions in twisted bilayer graphene}
	\date{\today}
	
	\author{D. O. Oriekhov}
	\affiliation{Kavli Institute of Nanoscience, Delft University of Technology, 2628 CJ Delft, the Netherlands}
	
	\author{T. T. Osterholt}
	\affiliation{Institute for Theoretical Physics, Utrecht University,
		Princetonplein 5, 3584CC Utrecht, The Netherlands}
	
	\author{R. A. Duine}
	\affiliation{Institute for Theoretical Physics, Utrecht University,
		Princetonplein 5, 3584CC Utrecht, The Netherlands}
	\affiliation{Department of Applied Physics, Eindhoven University of Technology,
			P.O. Box 513, 5600 MB Eindhoven, The Netherlands}
	
	\author{V. P. Gusynin}
	\affiliation{Bogolyubov Institute for Theoretical Physics, Kyiv, 03143, Ukraine}

	\begin{abstract}
		We study the competition between 
the RKKY quadratic and biquadratic spin-spin interactions of two magnetic impurities in twisted bilayer graphene away from
the magic angle. We apply the Bistritzer-MacDonald model of two graphene layers twisted with respect to each other by a small angle. By
reducing the model to the Dirac-type one with modified Fermi velocity, we derive expressions for the RKKY quadratic and biquadratic
spin interactions using perturbation theory for the free energy. The biquadratic interaction is suppressed by a larger power of the interaction
		constant and decreases faster with a the distance between impurities comparing to the quadratic one. Nevertheless, due to the different period of oscillations with impurity separation distance, chemical potential, twist angle and temperature, it is possible to fine-tune the system to the regime of dominating biquadratic interaction. The existence of such fine-tuned regime might provide a promising opportunity to observe non-conventional spin ordering.
	\end{abstract}
	\maketitle
	
	\section{Introduction}

	The study of exchange spin-spin interactions started from the pioneering work of Heisenberg on ferromagnetism \cite{Heisenberg1928}.
	One of the key questions appearing for all spin-spin interaction problems is the role of the surrounding medium.
	 A milestone in the studies of foundational principles of magnets was set with the discovery of the Ruderman-Kittel-Kasuya- Yosida (RKKY) interaction \cite{Ruderman1954,Kasuya1956,Yosida} which describes the exchange interaction between two magnetic
impurities induced by the conduction electrons of the medium. This usually appears as a leading order contribution from perturbation theory in the coupling constant between spin impurities and valence electrons of the underlying material. Integrating out the electronic degrees of freedom, one obtains the contribution of exchange interaction to the total free energy of the system. However, as was pointed out in a Ref. \cite{Kartsev2020NPJComput}, little is known about the next higher-order spin-spin interactions coming from the next terms in perturbation theory.
	The simplest non-Heisenberg coupling term of such kind that should be taken into account represents a biquadratic interaction: for two impurities with spins $\vec{S}_1$ and $\vec{S}_2$ it has the form  $(\vec{S}_1\vec{S}_2)^2$ for isotropic systems, whereas the standard RKKY term is $\vec{S}_1\vec{S}_2$.
	
		The model with biquadratic interaction was applied in Ref.\cite{Kartsev2020NPJComput} to describe magnetic phenomena in layered magnets
such as CrI${}_3$ and CrBr${}_3$. A number of candidate materials - such as $\mathrm{NiX}_2(\mathrm{X}=\mathrm{Cl}, \mathrm{Br}$ and $\mathrm{I})$
\cite{Ni2021} and iron-based superconductors \cite{Lima2021} - were studied where biquadratic spin couplings play a key role. In the theoretical
studies of effective bilinear-biquadratic models
of magnets having both RKKY quadratic and biquadratic interactions, it was found that unconventional magnetic order parameters could be formed:
quadrupole \cite{Kokorina2021,Kokorina2022}, spiral, stripe and tetrahedral orders \cite{Lima2021,Szasz2022}. In addition, a large biquadratic
interaction constant is expected to stabilize the ferromagnetic state in $\mathrm{NiX}_2(\mathrm{X}=\mathrm{Cl}, \mathrm{Br}$ and $\mathrm{I})$
\cite{Ni2021}.
	
	\begin{figure}
		\centering
		\includegraphics[scale=0.6]{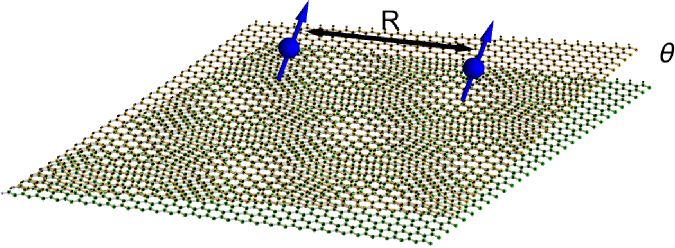}
		\caption{Schematic representation of the system considered in the paper: two impurities are placed on top of twisted bilayer graphene with layer rotation angle $\theta$. It is assumed that impurities are placed near individual atoms. The distance between impurities equals $R$.}
		\label{fig1:twisted-bilayer}
	\end{figure}

Usually, the biquadratic spin-spin interactions are added to phenomenological spin Hamiltonians to describe the stability and competition of
different phases of the system. And very rarely are such interactions derived from more microscopic theories. We will consider such a derivation
in this article, where we use the reduced low-energy Hamiltonian of the Bistritzer-Macdonald model \cite{Bistritzer2011} for twisted bilayer
graphene as a microscopic Hamiltonian.	The model system is presented in Fig.~\ref{fig1:twisted-bilayer}.
	The idea of twisting graphene layers to exploit an additional twist angle degree of freedom to vary interlayer electronic hopping
terms was first proposed in Ref.~\cite{LopesdosSantos2007} and tested experimentally in Ref.~\cite{LiAndrei2009Nature}. In the famous seminal
paper by Bistritzer and MacDonald \cite{Bistritzer2011} it was found that at specific (``magic'') angle a flat band is formed after
complete flattening of the Dirac cone. Later this prediction was confirmed in a number of experiments \cite{Cao2018,Cao2018second},
leading to the first observation of superconductivity in bilayer graphene without heavy doping.

In the present paper we focus on twist angles that are  away from the first magic angle, where the Dirac cone remains.
The corresponding model used in this study is obtained from the Bistritzer-Macdonald one. It reduces to an effective two-component Dirac Hamiltonian
with modified Fermi velocity depending on the rotation angle \cite{Lian2018,Catarina2019}. This allows for
complete analytic treatment of the RKKY interaction for all parameters \cite{Oriekhov2020}, and at zero temperature for the biquadratic interaction.
 The main finding of this paper is that the biquadratic interaction has different oscillations period with distance and doping
comparing  to the RKKY second order term at zero temperature. This implies a specific selection of twist angle and impurity
location, where the spin ordering would be predominantly determined by the biquadratic interaction.
	
	The paper is organized as follows: we start with introducing effective model of twisted bilayer graphene in
Sec.~\ref{sec:twisted-bilayer-model}. Next, using the free energy expression, we derive the contribution of biquadratic interaction
in terms of Green's functions of the free electron in Sec.~\ref{sec:general-expressions}, and obtain analytic expressions for interaction integrals in Sec.~\ref{sec:graphene-results-rkky}.
	Next, we present results for zero temperature case in Sec.~\ref{sec:zeto-T} and discuss the possibility of detecting biquadratic interaction at certain fine-tuned values of twist angle for given impurity positions. In Sec.\ref{sec:finite-temperature} we analyze
the role of finite temperature using numerically evaluated expressions for interaction integrals. We
present conclusions in Sec.~\ref{sec:conclusions}. Finally, appendix \ref{appendix} contains, as an example, the calculation of
the interaction integral to the second order of perturbation theory.
	
	\section{The effective model of twisted bilayer graphene}
	\label{sec:twisted-bilayer-model}
	 We start from the Bistritzer-MacDonald (BM) model \cite{Bistritzer2011}. It is obtained in the vicinity of
a single K-point by taking into account the fast decay of interlayer hopping parameter with distance. The BM model contains in total
eight spin-degenerate bands, and reduces to the following effective linearized model for the two lowest energy bands \cite{Bistritzer2011,Lian2018,Catarina2019} with effective Hamiltonian:
	\begin{align}\label{eq:Hamiltonian-main}
		{H}_{eff}(\mathbf{k})=\hbar v_F^* (\tau_x k_x+\xi \tau_y k_y),\, v_F^{*}=\left(\frac{1-3 \alpha^2}{1+6 \alpha^2}\right) v_F,
	\end{align}
	where $v_F^{*}$ is the effective Fermi velocity, Pauli matrices $\tau_x,\tau_y$
act on the layer degree of freedom in the spinor wave function, and $\xi$ is the valley index. The parameters in $v_F^*$ are defined through the
twist angle  $\theta$ and parameters of monolayer graphene as
	\begin{align}
		\alpha=w / \hbar v_F k_\theta,\quad k_\theta=8 \pi \sin (\theta / 2) / 3 a_0.
	\end{align}
	The numerical values used throughout the paper are: Fermi velocity of monolayer graphene $
		v_F= \sqrt{3} t a_0/2\hbar =9.3 \times 10^7 \mathrm{~cm}/\mathrm{s}$, lattice constant $ a_0=0.246 \mathrm{~nm}$, and the magnitude of
interlayer hopping parameter $w=110 \,\mathrm{meV}$. The approximation used to obtain Eq.\eqref{eq:Hamiltonian-main} imposes particle-hole symmetry.
For magic values of the angle, the effective Fermi velocity vanishes and the next order of the expansion in a wave vector should be taken into account. This results in the
appearance of van Hove singularities close to charge neutrality point \cite{Bistritzer2011,Yuan2019,Shankar2023}, for which the calculation of
spin-spin interactions in the perturbation theory would present a challenge due to the divergent density of
states. Thus, the calculations below always assume finite value of effective Fermi velocity. The model \eqref{eq:Hamiltonian-main} should work
decently well in the range of twist angles between $\theta\approx 0.8^\circ$ and $\theta=10^\circ$, where the lower
bound estimated from the middle between first and second magic angles being at $1.05^\circ$ and $0.5^\circ$, respectively; and the upper bound  was numerically estimated from applicability of Bloch's functions in Ref.\cite{Bistritzer2011}.
The energy range of applicability of the full BM model and of linearized two-band model is estimated to be up to $1$ eV from the charge-neutrality
point \cite{Bistritzer2011,Lian2018,Catarina2019}.
		
	The retarded Green's function of the model \eqref{eq:Hamiltonian-main} is given by
	\begin{align}
		G_0^R\left(\omega, \mathbf{k}, \xi\right)=\frac{\omega+\hbar v_F^{*}\left(\tau_x k_x+\xi \tau_y k_y\right)}{\left(\omega+i\epsilon\right)^2
-\left(\hbar v_F^{*}\right)^2 \mathbf{k}^2}.
\label{GF-momentum-space}
	\end{align}
	Using the results from monolayer graphene  with reduced Fermi velocity, the real space version
of the Green's function for a given valley index $\xi$ takes the form:
	\begin{align}\label{eq:GF-Hankel}
		&G_0^R\left(\mathbf{r}, \omega, \xi\right)= \frac{\omega}{4\left(\hbar v_F^{*}\right)^2}\times\nn
		&\left(\begin{array}{cc}-i  H_0^{(1)}(z) &  \xi e^{-i \xi \varphi} H_1^{(1)}(z) \\
			\xi e^{i \xi \varphi} H_1^{(1)}(z) & -i H_0^{(1)}(z)
		\end{array}\right),\quad z=\frac{|\mathbf{r}|(\omega+i \varepsilon)}{\hbar v_F^{*}} .
	\end{align}
	Here $H^{(1)}_i(z)$ is the Hankel function of the first kind and $\varphi$ is the polar angle measured from the $x$-axis.
	
	In the following sections we perform the calculation for only a single valley $\xi$ to extract the behavior of RKKY quadratic and biquadratic
interactions that is sensitive to Fermi velocity changes due to twist angle. Later we discuss the effects of taking into account two
valleys in the Moir\'{e} Brillouin zone.
	
	\section{Derivation of the RKKY quadratic and biquadratic interactions}
	\label{sec:general-expressions}
	For the purpose of deriving a general expression for the biquadratic interaction from perturbation theory, we start with the free energy
expressed through partition function as $F=-T \ln Z$:
	\begin{align}
		&Z
		=Z_0^{-1} \int \mathcal{D} \psi \mathcal{D} \psi^{\dagger} \times\nn
		&\exp \left\{ \int\limits_0^{1/T} \mathrm{d} \tau \int\mathrm{d}^2 x \psi^{\dagger}(\tau,x)
\left[-\frac{\partial}{\partial \tau}-H\right] \psi(\tau,x)\right\},
	\end{align}
where $T$ is the temperature (the Boltzmann constant $k_B$ is set equal to one) and fermion fields $\psi$ carry layer and
spin indices.

	The integration over fermionic fields $\psi$ takes into account low-energy electrons close to charge neutrality point.
	Here the Hamiltonian $H=H_0+V$ consists of two parts - kinetic part of free quasiparticles, $H_0$, in underlying material and the
interaction part, $V,$ which describes the coupling between magnetic impurities and  the
itinerant electrons of twisted bilayer graphene \cite{Saremi2007,Sherafati2011,Sherafati2011a,Kogan2011,Oriekhov2020}:
	 \begin{align}\label{eq:potential}
	 		V=-\lambda\left[\mathbf{S}_1 \cdot \mathbf{s} \delta\left(\mathbf{r}-\mathbf{R}_1\right) P_{l1}+\mathbf{S}_2 \cdot \mathbf{s} \delta\left(\mathbf{r}-\mathbf{R}_2\right) P_{l2}\right].
	 \end{align}
	 In this model spins $\mathbf{S}_1, \mathbf{S}_2$ of two impurities are assumed classical and the $\vec{s}=\vec{\sigma}/2$
operator stands for the electron spin in graphene expressed through the Pauli matrices, $\vec{R}_{1,2}$ are impurity positions and
 $P_{l1},P_{l2}$ are the projectors onto the layers where the respective impurities are placed (layer indices $l1,l2$ take
the values $1,2$). These projectors are diagonal matrices $P_{1}=\text{diag}(1,0)$ and $P_{2}=\text{diag}(0,1)$.
	The coupling constant $\lambda$ depends on the type of impurity placed on the graphene sheet.
In what follows we consider Co impurities bound to carbon atoms in monolayer graphene, in this case the coupling reaches the value $1 eV\cdot a_C$
where $a_C$ is the area per carbon atom $\simeq 2.62 \text{\AA}^2$ \cite{CaoFertig2019}.
	
	Since the action in partition function is quadratic in fermionic fields, we find the following result for the free energy:
	\begin{align}
		F&=-T \ln \frac{\operatorname{Det}\left[-\frac{\partial}{\partial \tau}-H\right]}{\operatorname{Det}\left[-\frac{\partial}{\partial \tau}-H_0\right]}\nn
		&=-T \operatorname{Tr} \ln \left( \left[-\frac{\partial}{\partial \tau}-H\right]\left[-\frac{\partial}{\partial \tau}-H_0\right]^{-1}\right).
	\end{align}
Here the $Tr$ operation includes matrix trace $tr$, summation over valleys and integration over coordinates. The last expression can be rewritten in terms of free particle Green's function via the substitution $-\frac{\partial}{\partial \tau}-H_0=G_0^{-1}$. This leads to the corresponding series expansion in powers of coupling constant $\lambda$:
	\begin{align}\label{eq:free-energy-decomposition}
		&F=-T \operatorname{Tr} \ln \hspace{-1.3pt}\left(1-V \hat{G}_0\right)=-T \operatorname{Tr}\hspace{-1.3pt}\left(-V \hat{G}_0-\frac{1}{2}
V \hat{G}_0 V \hat{G}_0\right.\nn
		&\left.-\frac{1}{3} V \hat{G}_0 V \hat{G}_0 V \hat{G}_0-\frac{1}{4} V \hat{G}_0 V \hat{G}_0 V \hat{G}_0 V \hat{G}_0-\ldots\right).
	\end{align}
In this expansion even powers of $V$ terms contain contributions to the RKKY quadratic interaction, and starting  the fourth order
additional biquadratic interactions appear (odd power terms vanish due to spin traces).

	Now we analyze second and fourth order contributions to the free energy. We evaluate the traces over spin matrix operators
taking into account that for graphene the Green's function is proportional to the unit matrix $\sigma_0\equiv \hat{1}$
in the real spin space. The combinatorial
coefficients from spin traces $\delta F^{\vec{S}}_n$  enter the full n-th order correction to free energy as $\delta F= \lambda^n \delta
F^{\vec{S}}_n I_{n}$, and $I_{n}$ contains integrals that depend on layer indices and distances between impurities.
For the spin traces we find:
	\begin{align}
		\delta F_{2}^{\vec{S}}&= \frac{1}{4}\tr[S_{1,i}\sigma_i  S_{2,j}\sigma_j ]=\frac{1}{2} \,\vec{S}_1\vec{S}_2\\
		\delta F_{3}^{\vec{S}}&= \frac{1}{8}\tr[S_{1,i}\sigma_i  S_{2,j}\sigma_j S_{1,k}\sigma_k]\nn
		&=\frac{1}{8}S_{1,i}S_{2,j}S_{1,k}\tr[\sigma_i (\delta_{j k} +i \varepsilon_{j k l} \sigma_{l})]=0\\
		\delta F_{4}^{\vec{S}} &= \frac{1}{16}\tr[(S_{1,i}\sigma_i)^4\hspace{-1.8pt}+\hspace{-1.8pt} S_{1,i}\sigma_i S_{2,j}\sigma_j
S_{1,k}\sigma_k S_{2,m}\sigma_m+\dots]\nn
		&= \frac{1}{8}[\vec{S}_1^2+\vec{S}_2^2+2\vec{S}_1\vec{S}_2]^2.
	\end{align}
	All odd contributions vanish due to the absence of odd power invariants composed of two spins that preserve rotational symmetry in the space.
  In the next calculations we also do not take into account energy shifts appearing from
 terms not containing dependence on the scalar product $\vec{S}_1\vec{S}_2$. Finally, from the expression for the fourth order, we extract
 constant terms, RKKY quadratic and biquadratic interaction:
	\begin{align}
		\delta F_{4}^{\vec{S}}= \frac{1}{8}[(\vec{S}_1^2+\vec{S}_2^2)^2+4(\vec{S}_1^2+\vec{S}_2^2)(\vec{S}_1\vec{S}_2) +4(\vec{S}_1\vec{S}_2)^2].
	\end{align}
	Having identified the orders and combinatorial coefficients of the leading contributions to the RKKY quadratic and biquadratic interactions,
we proceed with calculation of distance-dependent prefactors.
	
	\subsection{Expressions for distance-dependent prefactors}
	\label{sec:graphene-results-rkky}
	In the present subsection we extract the distance-dependent prefactors in both RKKY quadratic
	 and biquadratic interaction terms and write them in terms of integrals over frequency. The corresponding interaction strengths, which depend on a distance $R$ between two impurities, temperature and chemical potential $\mu$, are the prefactors of spin-dependent interaction terms:
\begin{align}
\hspace{-2mm}\delta F \hspace{-0.5mm}= \hspace{-0.5mm}J_{quad}(R,T,\mu)(\vec{S}_1\vec{S}_2)+J_{biq}(R,T,\mu)(\vec{S_1\vec{S}_2})^2.
\end{align}
  Substituting the real space  Green's function (\ref{eq:GF-Hankel})  into Eq.\eqref{eq:free-energy-decomposition} and performing the summation over  Matsubara frequencies by means of the well known formula (\ref{Matsubara-summation}), we arrive at the following expressions
	\begin{align}
\label{Jquad}
		&J_{quad}(R,T,\mu)=\frac{\lambda^2}{16(\hbar v_F^*)^4} I_{l1,l2}^{(2)}(R,T,\mu)\nn
		&+ (\vec{S}_1^2+\vec{S}_2^2)\frac{\lambda^4}{64(\hbar v_F^*)^8}I_{l1,l2}^{(4)}(R,T,\mu),\\
		&J_{biq}(R,T,\mu)=\frac{\lambda^4}{64(\hbar v_F^*)^8}I_{l1,l2}^{(4),biq}(R,T,\mu).
\label{Jbiq}
	\end{align}
	Here the indices $l1,l2$ denote the position of impurities in the spinor components of the Hamiltonian \eqref{eq:Hamiltonian-main} according
to projectors \eqref{eq:potential}. The summation over valley index was already performed in these expression, and resulted in an additional
factor $2$. In the lattice model the result could be further modified by the factor $1+\cos(\Delta \vec{K} \vec{R})$ with $\Delta \vec{K}=\vec{K}-\vec{K}'$. In what follows, for numerical calculations we take cobalt atoms with effective spin $S=3/2$ as impurities \cite{Fritz2013}. We remind that spins of magnetic impurities are considered as classical so that for cobalt $\vec{S}^2=9/4$.

The integrals defined above are expressed through the Hankel functions. In the case of the quadratic RKKY interaction we have to evaluate
the integrals:

	\begin{figure*}
	\centering
	\includegraphics[scale=0.41]{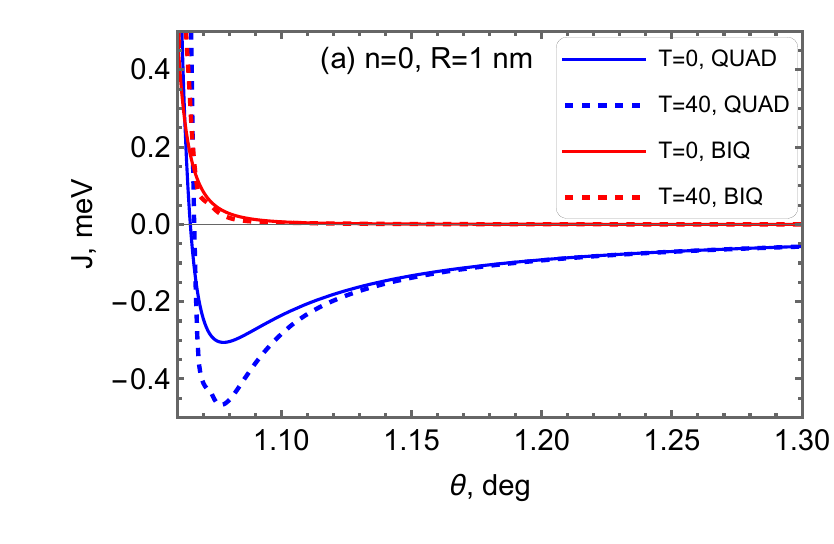}
	\includegraphics[scale=0.41]{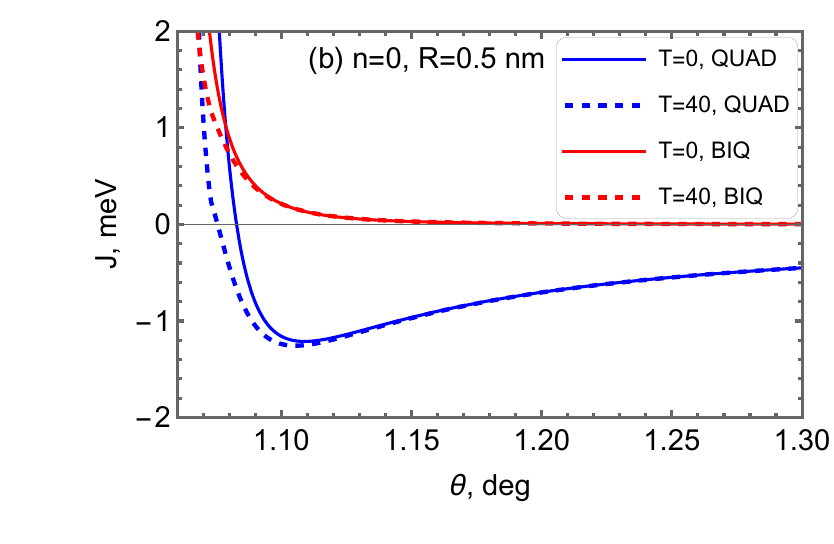}
	\includegraphics[scale=0.41]{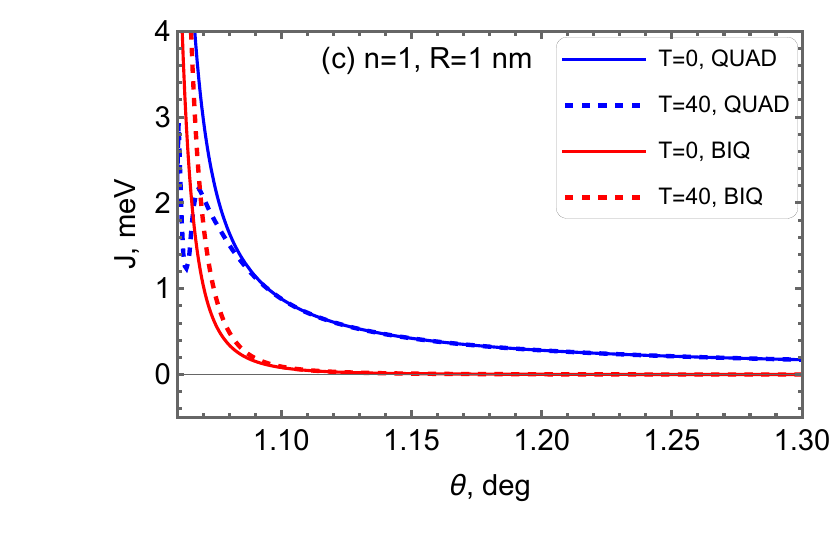}
	\caption{Comparison of interaction strengths dependencies on twist angle for $n=0$ (same layer) and $n=1$ (different layers). Interactions are evaluated via the integrals listed in Sec.\ref{sec:zeto-T} for zero temperature $T=0$, while the temperature corrections for $T=40 K$ are accounted with the integrals from Sec.\ref{sec:finite-temperature}. The interaction constant $\lambda=1$ eV $\cdot 2.62\text{\AA}^2$ and distance between impurities $R=1$ nm in panels (a) and (c), while $R=0.5 nm$ in panel (b). Chemical potential $\mu=1 $ meV lies within applicability range of the low-energy model for almost all angles. The strong enhancement of both interactions happens close to magic angle. For $n=0$ type interaction it is possible to find zero of quadratic interaction, while having nonzero biquadratic one. }
	\label{fig3:temperature-dependence}
\end{figure*}

	
	\begin{align}
		\label{eq:rkky-interaction-integral}
		& I_{l1,l2}^{(m)}(R,\mu,T)=\int_{-\infty}^{\infty} \frac{d \omega f_{l1,l2}^{(m)}(\omega)}{e^{\frac{\omega-\mu}{T}}+1},\\
		\label{eq:rkky-function-second-order}
		 &f_{l1,l2}^{(2)}(\omega)=\operatorname{Im}\left[(\omega+i \varepsilon)^2\left(H_n^{(1)}\left(\frac{(\omega+i \varepsilon) R}
{\hbar v_F^*}\right)\right)^2\right],\\
		\label{eq:rkky-fw-fourth-order}
		&f_{l1,l2}^{(4)}(\omega)=\operatorname{Im}\left[(\omega+i \varepsilon)^4\left(H_0^{(1)}\left(\frac{(\omega+i \varepsilon) R}
{\hbar v_F^*}\right)\right)^2\times\right.\nn
		&\left.\left(H_n^{(1)}\left(\frac{(\omega+i \varepsilon) R}{\hbar v_F^*}\right)\right)^2\right],
	\end{align}
	where $n=1-\delta_{l1,l2}$. Here $\mu$ is the chemical potential and $T$ is temperature measured in units of hopping parameter $t$ of monolayer graphene. In the case of biquadratic interaction, we find a different expression for $f(\omega)$ function:
	\begin{align}
		\label{eq:biquadratic-interaction-integral}
		&f_{l1,l2}^{biq}(\omega)= \operatorname{Im}\left[(\omega+i \varepsilon)^4\left(H_n^{(1)}\left(\frac{(\omega+i \varepsilon) R}
{\hbar v_F^*}\right)\right)^4\right],\nn &n=1-\delta_{l1,l2}.
	\end{align}
	One should note that the last expression is the same as Eq.\eqref{eq:rkky-fw-fourth-order} in the case of impurities being on the same
layer and sublattice $l1=l2$.
	
    Some integrals above, for example $I_{l1,l2}^{(2)}(R,\mu,T)$, can be evaluated using Mellin-Barnes transformation, which
    reduces them to a sums over various Meijer G-functions. This procedure was discussed in Ref.\cite{Oriekhov2020}.
    However, to describe qualitative behavior,
    the numerical evaluations at finite temperature are more appropriate. Thus, in the next sections we first analyze zero-temperature
    expressions and then discuss effects of finite temperature found by numerical evaluations.

	\section{Zero temperature limit}
	\label{sec:zeto-T}
	In the case of the RKKY  quadratic interaction at zero temperaturte the integral $_{l1,l2}^{(2)}(R,\mu,0)$ has the following analytic
form in terms of Meijer's G-function,
	\begin{align}
		&I_{l1,l2}^{(2)}(R, \mu, T=0)=\nn
		&=\left(\hspace{-0.5mm}\frac{\hbar v_F^*}{R}\hspace{-0.5mm}\right)^3\hspace{-0.5mm}\frac{1}{\sqrt{\pi}} G_{24}^{30}\hspace{-0.5mm}\left(\left(k_F R\right)^2\bigg|\bigg.\begin{array}{c}
			2,1\\
			0, \frac{3}{2}, \frac{3}{2}+n, \frac{3}{2}-n\end{array}\hspace{-1mm}\right).
\label{Meijer-representation}
	\end{align}
	Here Fermi wave vector $k_F$ is defined as $k_F=\mu/\hbar v_F^*$. For zero chemical potential we have
	\begin{align}
		G_{24}^{30}\left(0\bigg|\bigg.\begin{array}{c}
			2,1\\
			0, \frac{3}{2}, \frac{3}{2}+n, \frac{3}{2}-n\end{array}\right)=\frac{\left(4 n^2-1\right) \sqrt{\pi}}{8},
	\end{align}
hence
\begin{align}
I_{l1,l2}^{(2)}(R, 0, T=0)=\left(\frac{\hbar v_F^*}{R}\right)^3\frac{4 n^2-1}{8}.
\end{align}
To study asymptotical behaviour of our functions at large distances, $k_FR\gg1$, it is convenient to single out in the corresponding
zero-temperature integrals the parts that are independent of the chemical potential: $I(\mu)=I(0)+\int_0^\mu$. For the integrals
depending on $\mu$ we use an asymptotical expansion of Hankel's function (see chapter 10.17 in Ref.\cite{NIST:DLMF})
\begin{align}
H^{(1)}_\nu(z)\simeq\sqrt{\frac{2}{\pi z}}\,e^{i\phi}\sum\limits_{k=0}^\infty i^k\frac{a_k(\nu)}{z^k},
\end{align}
where
\begin{align*}
\phi=z-\frac{2\pi\nu+\pi}{4},\,a_k(\nu)=\frac{\Gamma\left(\frac{1}{2}-\nu+k\right)\Gamma\left(\frac{1}{2}+\nu+k\right)}
{\Gamma\left(\frac{1}{2}-\nu\right)\Gamma\left(\frac{1}{2}+\nu\right)(-2)^k k!}.
\end{align*}
Thus we find an asymptotical behaviour of oscillating part  at $k_FR\gg1$:
\begin{align}
&I_{l1,l2}^{(2)}(R, \mu, T=0)=\left(\frac{\hbar v_F^*}{R}\right)^3\frac{(-1)^{n+1}}{4\pi}\nn
&\times\left[4k_FR\sin(2k_FR)+(4n^2+1)\cos(2k_FR)\right].
\end{align}
The same asymptotical behaviour follows, of course, from Eq.(\ref{Meijer-representation}). A similar expression was obtained earlier in
studies of monolayer graphene \cite{Sherafati2011a,Klier2015} and pseudospin-1 system \cite{Oriekhov2020} where a corresponding $J$-integral described
the second-order interaction of impurities on sublattices.

	For the case of zero-temperature in fourth order term, we find the polynomial pre-factor $\left(\frac{\hbar v_F}{R}\right)^5$
for interaction integrals. Simple analytical expressions are obtained for the zero chemical potential by replacing integration
over the negative real axis with integration over the positive imaginary axis. Then using the well-known formula relating the Hankel
function of the imaginary argument to the modified Bessel function, we obtain
    \begin{align}\label{eq:integral-biq-4}
    	&I_{l1,l2}^{(4),biq}(R, 0, T=0) = \left(\frac{2}{\pi}\right)^4 \left(\frac{\hbar v_F^*}{R}\right)^5 \int\limits_{0}^{\infty}
    dz z^4 K_{n}^{4}(z) \nn
    	&=\left(\frac{2}{\pi}\right)^4 \left(\frac{\hbar v_F^*}{R}\right)^5\times \begin{cases}
    		0.046,& n=0\,\,(l1= l2) \\
    		0.561,& n=1\,\, (l1\neq l2).
    	\end{cases}\\
    \label{eq:integral-rkky-4}
    &I_{l1\neq l2}^{(4)}(R, 0, T=0)=\left(\frac{2}{\pi}\right)^4 \left(\frac{\hbar v_F^*}{R}\right)^5 \nn
    &\times\int\limits_{0}^{\infty} dz z^4 K_{0}^{2}(z)K_{1}^{2}(z)=\left(\frac{2}{\pi}\right)^4 \left(\frac{\hbar v_F^*}{R}\right)^5
    \times 0.132,\\
    &I_{l1=l2}^{(4),biq}(R, 0, T=0) = I_{l1= l2}^{(4)}(R, 0, T=0).
    \end{align}
    For the asymptotic behavior $k_FR\gg1$ we find
    \begin{align}
   &I^{(4)}_{l1,l2}(R,\mu,T=0)=\left(\frac{\hbar v_F^*}{R}\right)^5\frac{(-1)^{n+1}}{8\pi^2} \nn
   &\times\left[-8(k_FR)^2\cos(4k_F R)+8n^2(k_F R)\sin(4k_F R)\right.\nn
   &\left.+(3-6n^2+4n^4)\cos(4k_F R)\right],
\end{align}
\begin{align}
   &I^{(4),biq}_{l1,l2}(R,\mu,T=0)=-\left(\frac{\hbar v_F^*}{R}\right)^5\frac{1}{8\pi^2} \nn
   &\times\left[-8(k_F R)^2\cos(4k_F R)+16n^2(k_F R)\sin(4k_F R)\right.\nn
   &+\left.(3-12n^2+16n^4)\cos(4k_F R)\right].
    \end{align}

These results show that the biquadratic interaction $J_{biq}$ has generally a much faster decay with distance than the
    quadratic one, $J_{quad}$. This is connected, of course, with the presence of a contribution of order $\lambda^2$ in the interaction strength
    $J_{quad}$, see Eqs.(\ref{Jquad}), (\ref{Jbiq}). Thus, the long-range ordered phases defined
    by biquadratic interaction would be less stable with respect to perturbations. In the Fig.\ref{fig3:temperature-dependence} we compare the results
    of numerical evaluation for two different distances between impurities and a  chemical potential value $\mu=1$ meV with respect to to the change in twist angle value. Let us emphasize
    that the $J>0$ coupling for quadratic interaction is of ferromagnetic type, while $J<0$ is antiferromagnetic.

    As it is known for monolayer and bilayer graphene, the spin-spin interactions are generally weak. But, in the vicinity of magic angle, where
    the Dirac dispersion has a low effective Fermi velocity, interactions are strongly enhanced. That is clearly visible in all panels of Fig.\ref{fig3:temperature-dependence}. In addition, the oscillatory structure of interactions starts playing a role. The mathematical origin of
    these oscillations is related to quickly growing argument of Hankel functions under integrals when $v_F^{*}$ goes
    to zero. The grows of $1/v_F^{*}$ also sets a limit on applicability of perturbative expansion in Eq.\eqref{Jquad}. The dimensionless factor
    of the form $\lambda^2 / (2 \hbar v_F^{*} R)^2$ controls the ratio between first and second terms, and reaches a value of $1$ for angle $\theta=1.08^{\circ}$ and the distance between impurities $R=1$ nm.. At the same time, the small values of the integrals in Eq.\eqref{eq:integral-biq-4} and \eqref{eq:integral-rkky-4} further extend the applicability range, which is applied in Fig.\ref{fig3:temperature-dependence}. In the first two panels one can find a sequence of points at which the RKKY quadratic interaction passes zero, while biquadratic
    does not. This allows for turning off the RKKY quadratic interaction for twist angles below $1.1^{\circ}$ and short distances.
    Such a feature can be used to obtain a novel types of correlated states in twisted bilayer graphene
    by fine-tuning the distance between impurities.
     The dependence on chemical potential is weak in the applicability range of the model (see Appendix \ref{appendix-linearized}). At the same time, the distance between impurities allows for efficient control of the relative strength between interactions.

\begin{figure*}
	\centering
	\includegraphics[scale=0.5]{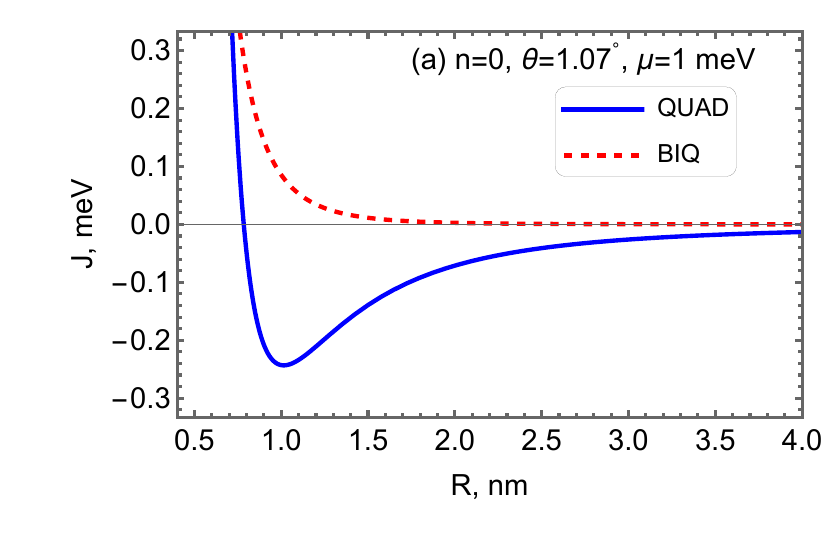}
	\includegraphics[scale=0.5]{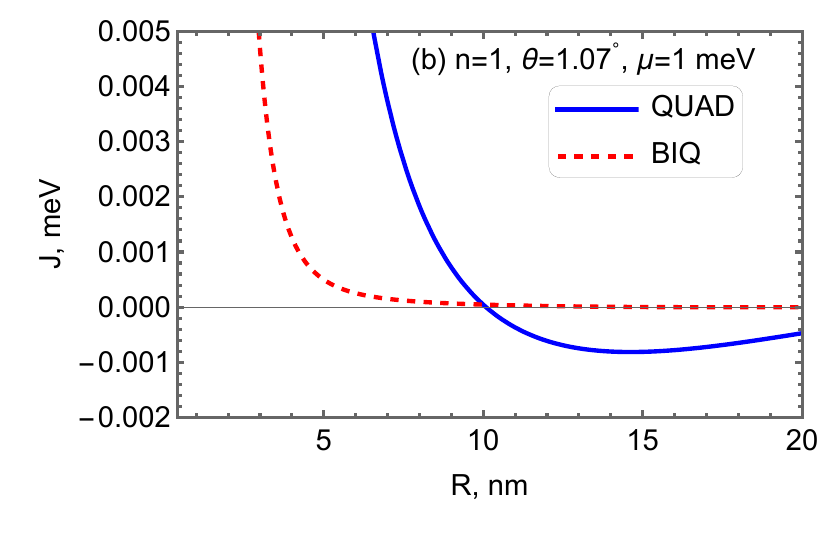}
	\caption{Distance dependence of interactions for zero temperature case and small chemical potential for the two example twist angles. The values of both interactions quickly decay with distance, but show larger relative value of biquadratic interaction than in monolayer graphene, see Fig.\ref{fig:monolayer-graphene}.}
	\label{fig:distance-dep-twisted}
\end{figure*}

\begin{figure}
	\centering
	\includegraphics[scale=0.5]{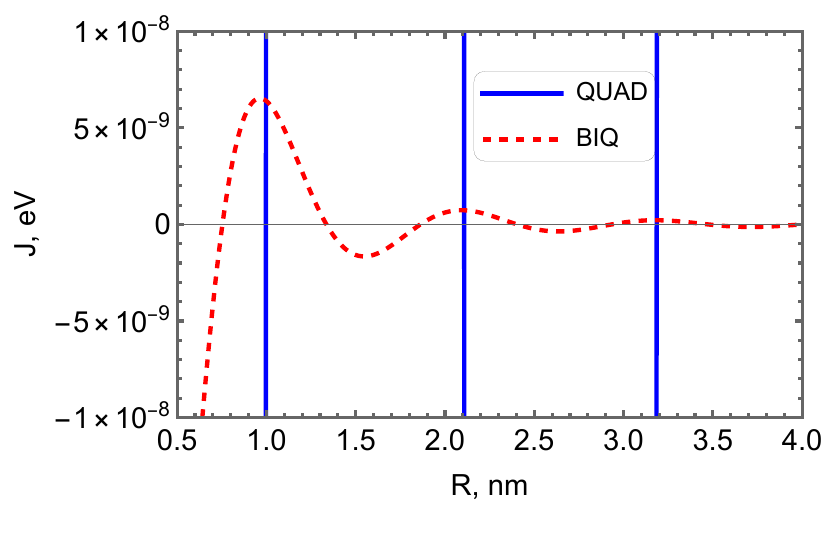}
	\caption{Distance dependence of interactions for monolayer graphene at chemical potential $\mu=0.9$ eV and the same sublattice position of impurities $n=0$. For the fine-tuned distance of impurities it is possible to achieve the dominant role of biquadratic interaction. However, the absolute values of both interactions for the distances of impurities more than $1$ nm are small. Even for quadratic interaction, the interaction values are of the order of one tens of meV \cite{Sherafati2011,Sherafati2011a}.}
	\label{fig:monolayer-graphene}
\end{figure}
In addition, we present an analysis of distance dependence of the quadratic and biquaratic interactions in Fig. \ref{fig:distance-dep-twisted}. Fig. \ref{fig:distance-dep-twisted} shows the optimal fine-tuned position of impurities to achieve the suppression of quadratic interaction for a fixed doping level of $1$ meV. This dependence should be compared with the same dependence for monolayer graphene, see Fig.\ref{fig:monolayer-graphene}. The results for monolayer graphene are obtained by setting the effective Fermi velocity to be the same as usual one, $v_F^{*}\to v_F$, in all expressions. The biquadratic interaction in the case of monolayer graphene is much weaker even for a very high doping level. Thus, we should point out, that while our  approach works for every Dirac-type system, the suitable observable results are expected to appear for a larger constant $\lambda$ of spin-spin interaction between impurities and band electrons.


\section{Temperature dependence}
\label{sec:finite-temperature}
	
	
	In this section we present numerical results for temperature-dependent case. The analysis contains both twist-angle dependence to estimate
	the possibility of observing effect at high temperatures and the chemical potential dependence at specific values of twist angle.
	
	The numerical integration is performed by dividing the integration interval into two parts, $[-\infty,0]$ and $[0,\infty]$ and changing the
sign of $\omega$ in the first case. Replacing the variables in Eq.\eqref{eq:rkky-interaction-integral}
with dimensionless ones, we find:
	\begin{align}
		\label{eq:T-z-decomposition}
		I_{l 1, l 2}^{(m)}(R, T, \mu)&=\left(\frac{\hbar v_F^*}{R}\right)^{m+1} \int_{0}^{\infty} d x f_{l1,l2}^{(m)}(x)\nn
		&\times\left(\frac{1}{z e^{x / T^{*}}+1}+\frac{z}{e^{x / T^{*}}+z}-1\right),
	\end{align}
	where  $T^{*}=\frac{T R}{\hbar v_F^*}, \quad z=e^{-\mu / T}$. The last term in Eq.\eqref{eq:T-z-decomposition} ($-1$ in the round brackets) describes the contribution at zero temperature and $\mu=0$, it diverges at the upper limit and thus requires a regularization. The appearance of divergences in separate terms of the perturbation series for the RKKY interaction was noticed long ago \cite{Vertogen1966,Bowen1968}  and is related to the local nature of the used RKKY interaction. The regularization could be done either by replacing the polynomial part of $f^m_{l1,l2}$ functions by $x^{\alpha-1}$ and further analytic continuation to needed values of $\alpha$ (see, for example, \cite{Oriekhov2020}), or implementing a finite frequency cutoff in the integral according to the energy range of applicability of the model \eqref{eq:Hamiltonian-main}.  In the last case it is important to use smooth cutoffs instead of a sharp one to obtain cutoff independent results in the long distance limit \cite{Saremi2007}. There is also the issue of the convergence of the entire perturbation series for the RKKY interaction raised in the recent work of Rusin and Zawadzki \cite{Rusin2020} which we discuss in more detail in the conclusions \ref{sec:conclusions}. 

	Taking into account the convergence subtlety in the expression above, the fully numerical calculation is more efficiently performed
via the following equivalent partition into temperature dependent and independent parts:
	 \begin{align}\label{eq:Integral-analogue-sommer}
	 	&I_{l 1, l 2}^{(m)}(R, T, \mu)=\left(\frac{\hbar v_F^*}{R}\right)^{m+1}\times\left[\int\limits_{-\infty}^{k_F R} d x f_{l1,l2}^{(m)}(x)\right.\nn
	 	&+\left. T^{*} \int\limits_0^{\infty} \frac{d x[f_{l1,l2}^{m}(k_F R+T^{*} x)-f_{l1,l2}^{m}(k_F R-T^{*} x)}{e^x+1}\right].
	 	\end{align}
 	In this form, the first term is known from the Sec.\ref{sec:zeto-T}. The last two terms represent a finite-temperature correction. In the brackets of the second integral, the function $f(\mu-T x)$ might contain jump at the point $x =\mu/T$. This feature is still integrable due to polynomial factors in all $f_{l1,l2}^{n}$ functions. However, it requires splitting of the integration interval at this point to ensure proper numerical convergence.

	Performing the evaluation for different angles starting from close to magic value $\theta=1.05^{\circ}$, we find the results presented in Fig.\ref{fig3:temperature-dependence}. The comparison with zero-temperature case shows that the $T=40$K contributes a correction in Eq.\eqref{eq:Integral-analogue-sommer} of the order of few percent for most of the twist angle values. But, it shifts the position of zero in quadratic interaction towards smaller angles.  
	The structure of oscillations close to first magic angle is altered, however the zeros of quadratic interaction do not match the zeros of biquadratic interaction. Thus, it is still possible to fine-tune the system to a regime when biquadratic interaction dominates.
	The varying value of the chemical potential has little influence on the results. The main contribution comes from the zero-doping integrals. Thus, we do not present separate plot of dependence on $\mu$.
	


	
	\section{Conclusions}
	\label{sec:conclusions}
	In the present paper we studied the twist angle dependence of the RKKY quadratic and biquadratic spin-spin interactions
 between two magnetic impurities mediated by itinerant electrons in twisted bilayer graphene away from the magic angle.
 General expressions for both interactions were derived from the free energy of the system with two impurities. The qualitative
 analysis shows that quadratic and
 biquadratic interactions have different oscillating terms, and thus there should exist regions in parameter
space of angle, distance between impurities, chemical potential and temperature, where the biquadratic interaction dominates.

	Using the analytic and numerical approaches, we show that in all cases it is possible to identify the angle and distance for which
the RKKY quadratic interaction vanishes, while the biquadratic one stays finite. This can lead to a formation of the new correlated phases
discussed in Refs.\cite{Szasz2022,Kokorina2021,Kokorina2022,Lima2021}, when a number of impurities are sparsely placed on top of graphene sheet.
 The oscillatory behavior of interactions close to magic angle shows the effect of band flattening on enhancement of both interactions with more
 fine-tuned competition between them.
	
Analyzing the results when angle approaches magic, we find the quick divergence of all interactions. From mathematical point of view, this is
a result of the trivial fact that the series expansion in $\lambda/\hbar v_F^{*} R$ loses its applicability
due to vanishing Fermi velocity. The physics  behind this is divergent density of states when the system approaches to the flat band.
In this regard, a more thorough study should be carried out along the lines of Refs. \cite{Oriekhov2020,Klinovaja2023} for the BM model.

Finally, we address the problem of convergence of entire RKKY perturbation series raised in the recent very interesting work of Rusin \& Zawadzki \cite{Rusin2020}. These authors obtained an exact RKKY Green’s function for electrons with parabolic isotropic dispersion at zero
temperature in three space dimensions. They got a criterion for the series convergence: the quantity $g_0=G_{free}(r=0,r’=0)$ must be finite, where $G_{free}(r,r')$ is the free-electron Green’s function.
Obtaining an exact Green’s function in the considered model with the effective Hamiltonian (1) is unsolved yet problem. Assuming that the
convergence criterion will be similar, we see that in our case it is not fulfilled due to integration by momenta of the free Green's function (\ref{GF-momentum-space}) up to infinity ($g_0$ is logarithmically divergent).  Certainly, there is a natural cutoff $k_c$ of the wave vector in
the model under consideration, related to the applicability domain of the model (see Appendix \ref{appendix-linearized}). Because of this, $g_0$
is finite and one can expect that the RKKY series of perturbation theory converges. 

The study of the RKKY interaction in this work indicates the ever-increasing role of such a control parameter as a twist angle in multilayer systems.
	As a future study, we expect the numerical analysis within the full Bistritzer-MacDonald model and similar effective models for
transition-metal dichalcogenides to be of great interest.
	
\section{Data and Code Availability}
The code for all numerical results presented in the paper can be found at the following  \cite{supplement-code} repository.

\section*{Acknowledgments}
 D.O.O. acknowledges the support by the Kavli Foundation. The work of T.T.O. and R.A.D. was supported by the Dutch Research Council (NWO) by the research programme Fluid Spintronics with Project No. 182.069 and by OCENW.XL21.XL21.058. The work of V.P.G. was supported by the Swiss NSF within the Ukrainian-Swiss Joint research project ``Transport and thermodynamic phenomena in low-dimensional materials with flat bands'' (grant No. IZURZ2\_224624).


\appendix
\section{Applicability range of the linearized model}
\label{appendix-linearized}
In this Appendix we discuss the applicability range of the linearized version of BM model - effective Dirac model near the band-touching point.

\begin{figure*}
	\centering
	\includegraphics[scale=0.5]{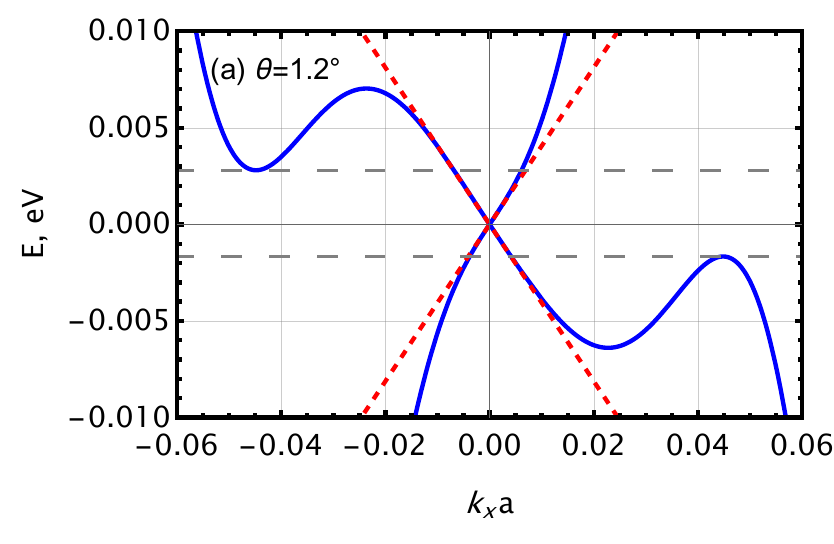}
	\includegraphics[scale=0.5]{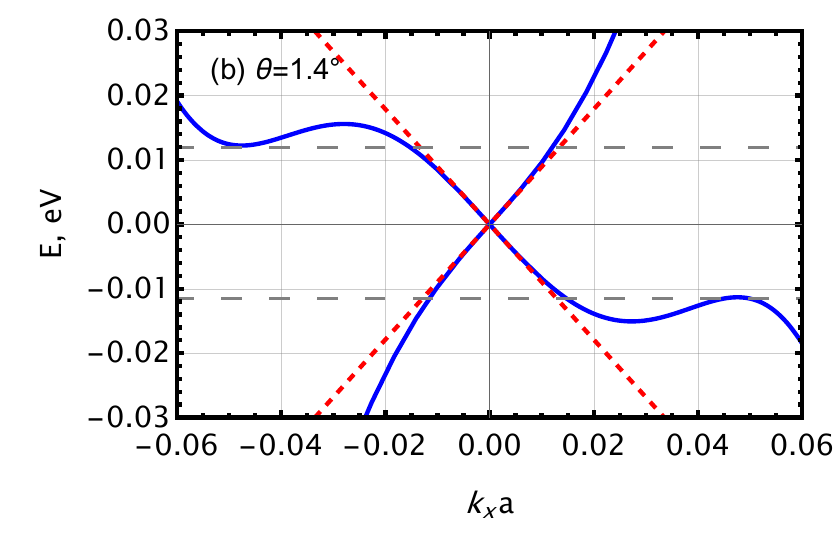}
	\caption{The comparison of effective Dirac model \eqref{eq:Hamiltonian-main} (red dashed lines) and Bistritzer-Macdonald model spectrum (solid lines) with band-touching point shifted to zero. Long-dashed lines show the chemical potential levels at which the linear Dirac model loses applicability independently of series expansion accuracy. This happens at local band minima, which are placed slightly away from K-points. Two different twist angles are shown.}
	\label{fig:applicability-range}
\end{figure*}

The main results are presented in Fig.\ref{fig:applicability-range}. They show that the calculations within Dirac model are limited to the narrow interval of chemical potentials. This interval is not symmetric at positive and negative sides, and shrinks towards zero when the twist angle approaches magic value. In the two panels presented in Fig.\ref{fig:applicability-range}, we show that the intervals are $\mu\in [ -1.65 , 2.8]\,meV$ for $\theta=1.2^{\circ}$ and $\mu\in [ -11.5 , 12]\,meV$ for $\theta=1.4^{\circ}$. This limitation of the linearized model in energy naturally transforms into the limiting allowed values of wave number $k_c a\approx 0.01$.

\section{Exact evaluation of the second-order contribution to free energy}
\label{appendix}
In this Appendix we demonstrate as an example of calculations,  the second order contribution of perturbation theory to the free energy defining the strength of the RKKY quadratic interaction. This contribution is given by the expression
\begin{widetext}
	\begin{align}
		& \delta F_2=\frac{\lambda^2 T}{2} \operatorname{tr} \int d \mathbf{r}_1 d \mathbf{r}_2 \int d \tau_1 d \tau_2\left[\left[\mathbf{S}_1
\cdot \frac{1}{2} \boldsymbol{\sigma} \delta\left(\mathbf{r}_1-\mathbf{R}_1\right) P_{l1}+\mathbf{S}_2 \cdot \frac{1}{2} \boldsymbol{\sigma} \delta\left(\mathbf{r}_1-\mathbf{R}_2\right) P_{l2}\right] \sigma_0 G_0\left(\mathbf{r}_1, \mathbf{r}_2 ; \tau_1-\tau_2\right) \times\right. \nn
		& \left.\times\left[\mathbf{S}_1 \cdot \frac{1}{2} \boldsymbol{\sigma} \delta\left(\mathbf{r}_2-\mathbf{R}_1\right) P_{l1}+\mathbf{S}_2 \cdot \frac{1}{2} \boldsymbol{\sigma} \delta\left(\mathbf{r}_2-\mathbf{R}_2\right) P_{l2}\right] \sigma_0 G_0\left(\mathbf{r}_2, \mathbf{r}_1 ; \tau_2-\tau_1\right)\right]
	\end{align}
\end{widetext}
Evaluating the integrals over delta-functions and performing trace operation over the spin matrices, we find:
\begin{align}
	&\delta F_{2}=\frac{\lambda^2}{4} \mathbf{S}_1 \mathbf{S}_2 \int_0^{1 / T} d \tau \operatorname{tr}\left[P_{l1} G_0\left(\mathbf{R}_1, \mathbf{R}_2 ; \tau\right) \right.\nn
	&\left.\times P_{l2} G_0\left(\mathbf{R}_2, \mathbf{R}_1 ;-\tau\right)+\left(l1\leftrightarrow l2, \mathbf{R}_1 \leftrightarrow  \mathbf{R}_2\right)\right].
\end{align}
The integration over $\tau$ can be equivalently rewritten as a sum over Matsubara frequencies
using the Fourier transform of imaginary-time Green function,
\begin{align}\label{eq:RKKY-graphene-matsubara-fourier}
	G_0(\tau)=T \sum_{n=-\infty}^\infty G_0\left(i \omega_{n}\right) e^{-i \omega_{n} \tau},\quad  \omega_{n}=(2 n+1) \pi T,
\end{align}
where $n$ is an integer. For $\delta F_{2}$ we get
\begin{align}
	\delta F_{2}=&\frac{\lambda^2}{2} \vec{S}_1 \vec{S}_2 T \sum_{n}\tr\left[P_{l{1}} G_0(\vec{R};i\omega_n+\mu)P_{l{2}}\right.
	\nonumber\\
	&\left. \emph{ }\times G_0(-\vec{R};i\omega_n+\mu)\right],
\end{align}
where $\vec{R}=\vec{R}_1-\vec{R}_2$ and we introduced the chemical potential $\mu$.
The sum over the Matsubara frequencies is performed by means of the formula
\begin{align}
	T\sum_{n}f(i\omega_n)=-\int\limits_{-\infty}^\infty\frac{d\omega}{\pi}n_F(\omega){\rm Im}f^R(\omega+i\epsilon),
	\label{Matsubara-summation}
\end{align}
where $n_F(\omega)=1/(\exp(\omega/T)+1)$ is the Fermi distribution function and superscript $R$ denotes retarded function.

To obtain results from the main text, one has
to substitute Green's function from Eq.\eqref{eq:GF-Hankel} with $\omega$ replaced by $\omega+\mu$.
 In considered case, the evaluation of traces over sublattice-layer degree of freedom results in a function presented in Eq.\eqref{eq:rkky-function-second-order}. For example, for the impurities placed on the same layer, we find
 \begin{align}
 	&\tr\left[\begin{pmatrix}
 		1 & 0\\
 		0 & 0
 	\end{pmatrix} \frac{\omega+\mu}{4\left(\hbar v_F\right)^2}\left(\begin{array}{cc}
 	-i H_0^{(1)}(z) & \xi e^{-i \xi \varphi} H_1^{(1)}(z) \\
 	\xi e^{i \xi \varphi} H_1^{(1)}(z) & -i H_0^{(1)}(z)
 	\end{array}\right)\right.\nn
 	&\left.\times \begin{pmatrix}
 	1 & 0\\
 	0 & 0
 	\end{pmatrix}\frac{\omega+\mu}{4\left(\hbar v_F\right)^2}\right.\nn
 &\times\left.\left(\begin{array}{cc}
 	-i H_0^{(1)}(z) & \xi e^{-i \xi (\pi+\varphi)} H_1^{(1)}(z) \\
 	\xi e^{i \xi (\pi+\varphi)} H_1^{(1)}(z) & -i H_0^{(1)}(z)
 	\end{array}\right) \right]\nn
 	&=\frac{(\omega+\mu)^2}{16\left(\hbar v_F\right)^4}\left(H_{0}^{1}(z)\right)^2,\quad z=\frac{|\vec{R}|(\omega+\mu)}{\hbar v_F^{*}}.
 \end{align}

Similarly, we can evaluate the fourth-order correction to the free energy, and thus come to the main expressions in the section
\ref{sec:graphene-results-rkky}.

\section{The convergence of series expansion}
\label{sec:convergence-section}
In this Appendix we analyze the convergence of series expansion by evaluating the relative coupling constant. This relative constant is defined as $\lambda^2 / (2 \hbar v_F^{*}R)^2$ and fixes the factor by which the next terms after biquadratic interaction are suppressed. At the same time, it does not take into account the values of the zero doping integrals itself, which could be smaller than 1.
\begin{figure}
	\centering
	\includegraphics[scale=0.5]{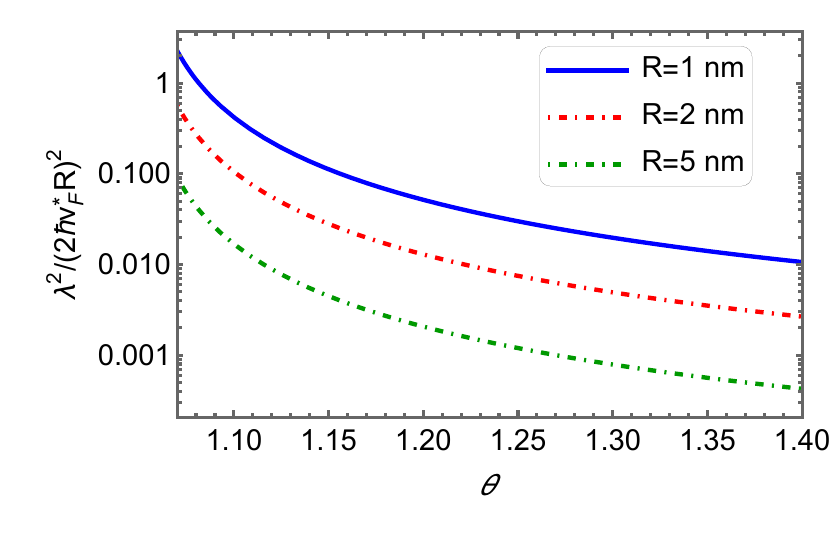}
	\caption{The angle dependence of dimensionless coupling constant that controls the ratio between RKKY and biquadratic interactions. }
	\label{fig:relative-strength-comparison}
\end{figure}


\bibliography{biquadratic_bib}

\begin{thebibliography}{33}%
\makeatletter
\providecommand \@ifxundefined [1]{%
 \@ifx{#1\undefined}
}%
\providecommand \@ifnum [1]{%
 \ifnum #1\expandafter \@firstoftwo
 \else \expandafter \@secondoftwo
 \fi
}%
\providecommand \@ifx [1]{%
 \ifx #1\expandafter \@firstoftwo
 \else \expandafter \@secondoftwo
 \fi
}%
\providecommand \natexlab [1]{#1}%
\providecommand \enquote  [1]{``#1''}%
\providecommand \bibnamefont  [1]{#1}%
\providecommand \bibfnamefont [1]{#1}%
\providecommand \citenamefont [1]{#1}%
\providecommand \href@noop [0]{\@secondoftwo}%
\providecommand \href [0]{\begingroup \@sanitize@url \@href}%
\providecommand \@href[1]{\@@startlink{#1}\@@href}%
\providecommand \@@href[1]{\endgroup#1\@@endlink}%
\providecommand \@sanitize@url [0]{\catcode `\\12\catcode `\$12\catcode
  `\&12\catcode `\#12\catcode `\^12\catcode `\_12\catcode `\%12\relax}%
\providecommand \@@startlink[1]{}%
\providecommand \@@endlink[0]{}%
\providecommand \url  [0]{\begingroup\@sanitize@url \@url }%
\providecommand \@url [1]{\endgroup\@href {#1}{\urlprefix }}%
\providecommand \urlprefix  [0]{URL }%
\providecommand \Eprint [0]{\href }%
\providecommand \doibase [0]{https://doi.org/}%
\providecommand \selectlanguage [0]{\@gobble}%
\providecommand \bibinfo  [0]{\@secondoftwo}%
\providecommand \bibfield  [0]{\@secondoftwo}%
\providecommand \translation [1]{[#1]}%
\providecommand \BibitemOpen [0]{}%
\providecommand \bibitemStop [0]{}%
\providecommand \bibitemNoStop [0]{.\EOS\space}%
\providecommand \EOS [0]{\spacefactor3000\relax}%
\providecommand \BibitemShut  [1]{\csname bibitem#1\endcsname}%
\let\auto@bib@innerbib\@empty
\bibitem [{\citenamefont {Heisenberg}(1928)}]{Heisenberg1928}%
  \BibitemOpen
  \bibfield  {author} {\bibinfo {author} {\bibfnamefont {W.}~\bibnamefont
  {Heisenberg}},\ }\bibfield  {title} {\bibinfo {title} {Zur theorie des
  ferromagnetismus},\ }\href {https://doi.org/10.1007/bf01328601} {\bibfield
  {journal} {\bibinfo  {journal} {Zeitschrift f\"{u}r Physik}\ }\textbf
  {\bibinfo {volume} {49}},\ \bibinfo {pages} {619} (\bibinfo {year}
  {1928})}\BibitemShut {NoStop}%
\bibitem [{\citenamefont {Ruderman}\ and\ \citenamefont
  {Kittel}(1954)}]{Ruderman1954}%
  \BibitemOpen
  \bibfield  {author} {\bibinfo {author} {\bibfnamefont {M.~A.}\ \bibnamefont
  {Ruderman}}\ and\ \bibinfo {author} {\bibfnamefont {C.}~\bibnamefont
  {Kittel}},\ }\bibfield  {title} {\bibinfo {title} {Indirect exchange coupling
  of nuclear magnetic moments by conduction electrons},\ }\href
  {https://doi.org/10.1103/PhysRev.96.99} {\bibfield  {journal} {\bibinfo
  {journal} {Phys. Rev.}\ }\textbf {\bibinfo {volume} {96}},\ \bibinfo {pages}
  {99} (\bibinfo {year} {1954})}\BibitemShut {NoStop}%
\bibitem [{\citenamefont {Kasuya}(1956)}]{Kasuya1956}%
  \BibitemOpen
  \bibfield  {author} {\bibinfo {author} {\bibfnamefont {T.}~\bibnamefont
  {Kasuya}},\ }\bibfield  {title} {\bibinfo {title} {A theory of metallic
  ferro- and antiferromagnetism on {Z}ener’s model},\ }\href
  {https://doi.org/10.1143/ptp.16.45} {\bibfield  {journal} {\bibinfo
  {journal} {Progress of Theoretical Physics}\ }\textbf {\bibinfo {volume}
  {16}},\ \bibinfo {pages} {45} (\bibinfo {year} {1956})}\BibitemShut {NoStop}%
\bibitem [{\citenamefont {Yosida}(1957)}]{Yosida}%
  \BibitemOpen
  \bibfield  {author} {\bibinfo {author} {\bibfnamefont {K.}~\bibnamefont
  {Yosida}},\ }\bibfield  {title} {\bibinfo {title} {Magnetic properties of
  {C}u-{M}n alloys},\ }\href {https://doi.org/10.1103/PhysRev.106.893}
  {\bibfield  {journal} {\bibinfo  {journal} {Phys. Rev.}\ }\textbf {\bibinfo
  {volume} {106}},\ \bibinfo {pages} {893} (\bibinfo {year}
  {1957})}\BibitemShut {NoStop}%
\bibitem [{\citenamefont {Kartsev}\ \emph {et~al.}(2020)\citenamefont
  {Kartsev}, \citenamefont {Augustin}, \citenamefont {Evans}, \citenamefont
  {Novoselov},\ and\ \citenamefont {Santos}}]{Kartsev2020NPJComput}%
  \BibitemOpen
  \bibfield  {author} {\bibinfo {author} {\bibfnamefont {A.}~\bibnamefont
  {Kartsev}}, \bibinfo {author} {\bibfnamefont {M.}~\bibnamefont {Augustin}},
  \bibinfo {author} {\bibfnamefont {R.~F.~L.}\ \bibnamefont {Evans}}, \bibinfo
  {author} {\bibfnamefont {K.~S.}\ \bibnamefont {Novoselov}},\ and\ \bibinfo
  {author} {\bibfnamefont {E.~J.~G.}\ \bibnamefont {Santos}},\ }\bibfield
  {title} {\bibinfo {title} {Biquadratic exchange interactions in
  two-dimensional magnets},\ }\href
  {https://doi.org/10.1038/s41524-020-00416-1} {\bibfield  {journal} {\bibinfo
  {journal} {npj Computational Materials}\ }\textbf {\bibinfo {volume} {6}},\
  \bibinfo {pages} {150} (\bibinfo {year} {2020})}\BibitemShut {NoStop}%
\bibitem [{\citenamefont {Ni}\ \emph {et~al.}(2021)\citenamefont {Ni},
  \citenamefont {Li}, \citenamefont {Amoroso}, \citenamefont {He},
  \citenamefont {Feng}, \citenamefont {Kan}, \citenamefont {Picozzi},\ and\
  \citenamefont {Xiang}}]{Ni2021}%
  \BibitemOpen
  \bibfield  {author} {\bibinfo {author} {\bibfnamefont {J.~Y.}\ \bibnamefont
  {Ni}}, \bibinfo {author} {\bibfnamefont {X.~Y.}\ \bibnamefont {Li}}, \bibinfo
  {author} {\bibfnamefont {D.}~\bibnamefont {Amoroso}}, \bibinfo {author}
  {\bibfnamefont {X.}~\bibnamefont {He}}, \bibinfo {author} {\bibfnamefont
  {J.~S.}\ \bibnamefont {Feng}}, \bibinfo {author} {\bibfnamefont {E.~J.}\
  \bibnamefont {Kan}}, \bibinfo {author} {\bibfnamefont {S.}~\bibnamefont
  {Picozzi}},\ and\ \bibinfo {author} {\bibfnamefont {H.~J.}\ \bibnamefont
  {Xiang}},\ }\bibfield  {title} {\bibinfo {title} {Giant biquadratic exchange
  in 2d magnets and its role in stabilizing ferromagnetism of {N}i{C}l${}_2$
  monolayers},\ }\href {https://doi.org/10.1103/physrevlett.127.247204}
  {\bibfield  {journal} {\bibinfo  {journal} {Physical Review Letters}\
  }\textbf {\bibinfo {volume} {127}},\ \bibinfo {pages} {247204} (\bibinfo
  {year} {2021})}\BibitemShut {NoStop}%
\bibitem [{\citenamefont {Lima}(2021)}]{Lima2021}%
  \BibitemOpen
  \bibfield  {author} {\bibinfo {author} {\bibfnamefont {L.~S.}\ \bibnamefont
  {Lima}},\ }\bibfield  {title} {\bibinfo {title} {Quantum correlation in the
  bilinear–biquadratic model for iron-based superconductors},\ }\href
  {https://doi.org/10.1140/epjp/s13360-021-01779-0} {\bibfield  {journal}
  {\bibinfo  {journal} {The European Physical Journal Plus}\ }\textbf {\bibinfo
  {volume} {136}},\ \bibinfo {pages} {789} (\bibinfo {year}
  {2021})}\BibitemShut {NoStop}%
\bibitem [{\citenamefont {Kokorina}\ and\ \citenamefont
  {Medvedev}(2021)}]{Kokorina2021}%
  \BibitemOpen
  \bibfield  {author} {\bibinfo {author} {\bibfnamefont {E.~E.}\ \bibnamefont
  {Kokorina}}\ and\ \bibinfo {author} {\bibfnamefont {M.~V.}\ \bibnamefont
  {Medvedev}},\ }\bibfield  {title} {\bibinfo {title} {Magnetocaloric effect in
  a first-order phase transition in a ferromagnet with biquadratic exchange},\
  }\href {https://doi.org/10.1134/s0031918x21110089} {\bibfield  {journal}
  {\bibinfo  {journal} {Physics of Metals and Metallography}\ }\textbf
  {\bibinfo {volume} {122}},\ \bibinfo {pages} {1045} (\bibinfo {year}
  {2021})}\BibitemShut {NoStop}%
\bibitem [{\citenamefont {Kokorina}\ and\ \citenamefont
  {Medvedev}(2022)}]{Kokorina2022}%
  \BibitemOpen
  \bibfield  {author} {\bibinfo {author} {\bibfnamefont {E.~E.}\ \bibnamefont
  {Kokorina}}\ and\ \bibinfo {author} {\bibfnamefont {M.~V.}\ \bibnamefont
  {Medvedev}},\ }\bibfield  {title} {\bibinfo {title} {Quadrupole ordering and
  inverse magnetocaloric effect in a magnet with biquadratic exchange and spin
  {S} = 1},\ }\href {https://doi.org/10.1134/s0031918x2209006x} {\bibfield
  {journal} {\bibinfo  {journal} {Physics of Metals and Metallography}\
  }\textbf {\bibinfo {volume} {123}},\ \bibinfo {pages} {878} (\bibinfo {year}
  {2022})}\BibitemShut {NoStop}%
\bibitem [{\citenamefont {Szasz}\ \emph {et~al.}(2022)\citenamefont {Szasz},
  \citenamefont {Wang},\ and\ \citenamefont {He}}]{Szasz2022}%
  \BibitemOpen
  \bibfield  {author} {\bibinfo {author} {\bibfnamefont {A.}~\bibnamefont
  {Szasz}}, \bibinfo {author} {\bibfnamefont {C.}~\bibnamefont {Wang}},\ and\
  \bibinfo {author} {\bibfnamefont {Y.-C.}\ \bibnamefont {He}},\ }\bibfield
  {title} {\bibinfo {title} {Phase diagram of a bilinear-biquadratic spin-1
  model on the triangular lattice from density matrix renormalization group
  simulations},\ }\href {https://doi.org/10.1103/physrevb.106.115103}
  {\bibfield  {journal} {\bibinfo  {journal} {Physical Review B}\ }\textbf
  {\bibinfo {volume} {106}},\ \bibinfo {pages} {115103} (\bibinfo {year}
  {2022})}\BibitemShut {NoStop}%
\bibitem [{\citenamefont {Bistritzer}\ and\ \citenamefont
  {MacDonald}(2011)}]{Bistritzer2011}%
  \BibitemOpen
  \bibfield  {author} {\bibinfo {author} {\bibfnamefont {R.}~\bibnamefont
  {Bistritzer}}\ and\ \bibinfo {author} {\bibfnamefont {A.~H.}\ \bibnamefont
  {MacDonald}},\ }\bibfield  {title} {\bibinfo {title} {Moir\'{e} bands in
  twisted double-layer graphene},\ }\href
  {https://doi.org/10.1073/pnas.1108174108} {\bibfield  {journal} {\bibinfo
  {journal} {Proceedings of the National Academy of Sciences}\ }\textbf
  {\bibinfo {volume} {108}},\ \bibinfo {pages} {12233} (\bibinfo {year}
  {2011})}\BibitemShut {NoStop}%
\bibitem [{\citenamefont {Lopes dos Santos}\ \emph
  {et~al.}(2007)\citenamefont {Lopes dos Santos}, \citenamefont {Peres},\
  and\ \citenamefont {Castro Neto}}]{LopesdosSantos2007}%
  \BibitemOpen
  \bibfield  {author} {\bibinfo {author} {\bibfnamefont {J.~M.~B.}\
  \bibnamefont {Lopes dos Santos}}, \bibinfo {author} {\bibfnamefont
  {N.~M.~R.}\ \bibnamefont {Peres}},\ and\ \bibinfo {author} {\bibfnamefont
  {A.~H.}\ \bibnamefont {Castro Neto}},\ }\bibfield  {title} {\bibinfo {title}
  {Graphene bilayer with a twist: electronic structure},\ }\href
  {https://doi.org/10.1103/physrevlett.99.256802} {\bibfield  {journal}
  {\bibinfo  {journal} {Physical Review Letters}\ }\textbf {\bibinfo {volume}
  {99}},\ \bibinfo {pages} {256802} (\bibinfo {year} {2007})}\BibitemShut
  {NoStop}%
\bibitem [{\citenamefont {Li}\ \emph {et~al.}(2009)\citenamefont {Li},
  \citenamefont {Luican}, \citenamefont {Lopes~dos Santos}, \citenamefont
  {Castro~Neto}, \citenamefont {Reina}, \citenamefont {Kong},\ and\
  \citenamefont {Andrei}}]{LiAndrei2009Nature}%
  \BibitemOpen
  \bibfield  {author} {\bibinfo {author} {\bibfnamefont {G.}~\bibnamefont
  {Li}}, \bibinfo {author} {\bibfnamefont {A.}~\bibnamefont {Luican}}, \bibinfo
  {author} {\bibfnamefont {J.~M.~B.}\ \bibnamefont {Lopes~dos Santos}},
  \bibinfo {author} {\bibfnamefont {A.~H.}\ \bibnamefont {Castro~Neto}},
  \bibinfo {author} {\bibfnamefont {A.}~\bibnamefont {Reina}}, \bibinfo
  {author} {\bibfnamefont {J.}~\bibnamefont {Kong}},\ and\ \bibinfo {author}
  {\bibfnamefont {E.~Y.}\ \bibnamefont {Andrei}},\ }\bibfield  {title}
  {\bibinfo {title} {Observation of {V}an {H}ove singularities in twisted
  graphene layers},\ }\href {https://doi.org/10.1038/nphys1463} {\bibfield
  {journal} {\bibinfo  {journal} {Nature Physics}\ }\textbf {\bibinfo {volume}
  {6}},\ \bibinfo {pages} {109} (\bibinfo {year} {2009})}\BibitemShut {NoStop}%
\bibitem [{\citenamefont {Cao}\ \emph {et~al.}(2018{\natexlab{a}})\citenamefont
  {Cao}, \citenamefont {Fatemi}, \citenamefont {Fang}, \citenamefont
  {Watanabe}, \citenamefont {Taniguchi}, \citenamefont {Kaxiras},\ and\
  \citenamefont {Jarillo-Herrero}}]{Cao2018}%
  \BibitemOpen
  \bibfield  {author} {\bibinfo {author} {\bibfnamefont {Y.}~\bibnamefont
  {Cao}}, \bibinfo {author} {\bibfnamefont {V.}~\bibnamefont {Fatemi}},
  \bibinfo {author} {\bibfnamefont {S.}~\bibnamefont {Fang}}, \bibinfo {author}
  {\bibfnamefont {K.}~\bibnamefont {Watanabe}}, \bibinfo {author}
  {\bibfnamefont {T.}~\bibnamefont {Taniguchi}}, \bibinfo {author}
  {\bibfnamefont {E.}~\bibnamefont {Kaxiras}},\ and\ \bibinfo {author}
  {\bibfnamefont {P.}~\bibnamefont {Jarillo-Herrero}},\ }\bibfield  {title}
  {\bibinfo {title} {Unconventional superconductivity in magic-angle graphene
  superlattices},\ }\href {https://doi.org/10.1038/nature26160} {\bibfield
  {journal} {\bibinfo  {journal} {Nature}\ }\textbf {\bibinfo {volume} {556}},\
  \bibinfo {pages} {43} (\bibinfo {year} {2018}{\natexlab{a}})}\BibitemShut
  {NoStop}%
\bibitem [{\citenamefont {Cao}\ \emph {et~al.}(2018{\natexlab{b}})\citenamefont
  {Cao}, \citenamefont {Fatemi}, \citenamefont {Demir}, \citenamefont {Fang},
  \citenamefont {Tomarken}, \citenamefont {Luo}, \citenamefont
  {Sanchez-Yamagishi}, \citenamefont {Watanabe}, \citenamefont {Taniguchi},
  \citenamefont {Kaxiras}, \citenamefont {Ashoori},\ and\ \citenamefont
  {Jarillo-Herrero}}]{Cao2018second}%
  \BibitemOpen
  \bibfield  {author} {\bibinfo {author} {\bibfnamefont {Y.}~\bibnamefont
  {Cao}}, \bibinfo {author} {\bibfnamefont {V.}~\bibnamefont {Fatemi}},
  \bibinfo {author} {\bibfnamefont {A.}~\bibnamefont {Demir}}, \bibinfo
  {author} {\bibfnamefont {S.}~\bibnamefont {Fang}}, \bibinfo {author}
  {\bibfnamefont {S.~L.}\ \bibnamefont {Tomarken}}, \bibinfo {author}
  {\bibfnamefont {J.~Y.}\ \bibnamefont {Luo}}, \bibinfo {author} {\bibfnamefont
  {J.~D.}\ \bibnamefont {Sanchez-Yamagishi}}, \bibinfo {author} {\bibfnamefont
  {K.}~\bibnamefont {Watanabe}}, \bibinfo {author} {\bibfnamefont
  {T.}~\bibnamefont {Taniguchi}}, \bibinfo {author} {\bibfnamefont
  {E.}~\bibnamefont {Kaxiras}}, \bibinfo {author} {\bibfnamefont {R.~C.}\
  \bibnamefont {Ashoori}},\ and\ \bibinfo {author} {\bibfnamefont
  {P.}~\bibnamefont {Jarillo-Herrero}},\ }\bibfield  {title} {\bibinfo {title}
  {Correlated insulator behaviour at half-filling in magic-angle graphene
  superlattices},\ }\href {https://doi.org/10.1038/nature26154} {\bibfield
  {journal} {\bibinfo  {journal} {Nature}\ }\textbf {\bibinfo {volume} {556}},\
  \bibinfo {pages} {80} (\bibinfo {year} {2018}{\natexlab{b}})}\BibitemShut
  {NoStop}%
\bibitem [{\citenamefont {Lian}\ \emph {et~al.}(2019)\citenamefont {Lian},
  \citenamefont {Wang},\ and\ \citenamefont {Bernevig}}]{Lian2018}%
  \BibitemOpen
  \bibfield  {author} {\bibinfo {author} {\bibfnamefont {B.}~\bibnamefont
  {Lian}}, \bibinfo {author} {\bibfnamefont {Z.}~\bibnamefont {Wang}},\ and\
  \bibinfo {author} {\bibfnamefont {B.~A.}\ \bibnamefont {Bernevig}},\
  }\bibfield  {title} {\bibinfo {title} {Twisted bilayer graphene: a phonon
  driven superconductor},\ }\href
  {https://doi.org/10.1103/physrevlett.122.257002} {\bibfield  {journal}
  {\bibinfo  {journal} {Phys. Rev. Lett.}\ }\textbf {\bibinfo {volume} {122}},\
  \bibinfo {pages} {257002} (\bibinfo {year} {2019})}\BibitemShut {NoStop}%
\bibitem [{\citenamefont {Catarina}\ \emph {et~al.}(2019)\citenamefont
  {Catarina}, \citenamefont {Amorim}, \citenamefont {Castro}, \citenamefont
  {Lopes},\ and\ \citenamefont {Peres}}]{Catarina2019}%
  \BibitemOpen
  \bibfield  {author} {\bibinfo {author} {\bibfnamefont {G.}~\bibnamefont
  {Catarina}}, \bibinfo {author} {\bibfnamefont {B.}~\bibnamefont {Amorim}},
  \bibinfo {author} {\bibfnamefont {E.~V.}\ \bibnamefont {Castro}}, \bibinfo
  {author} {\bibfnamefont {J.~M. V.~P.}\ \bibnamefont {Lopes}},\ and\ \bibinfo
  {author} {\bibfnamefont {N.~M.~R.}\ \bibnamefont {Peres}},\ }\bibfield
  {title} {\bibinfo {title} {Twisted bilayer graphene: low-energy physics,
  electronic and optical properties},\ }\href
  {https://doi.org/10.1002/9781119468455.ch44} {\bibfield  {journal} {\bibinfo
  {journal} {Handbook of Graphene, Volume 3: Graphene-like 2D Materials, Edited
  by Mei Zhang (John Wiley, 2019), Chap. 6}\ ,\ \bibinfo {pages} {177}}
  (\bibinfo {year} {2019})}\BibitemShut {NoStop}%
\bibitem [{\citenamefont {Oriekhov}\ and\ \citenamefont
  {Gusynin}(2020)}]{Oriekhov2020}%
  \BibitemOpen
  \bibfield  {author} {\bibinfo {author} {\bibfnamefont {D.~O.}\ \bibnamefont
  {Oriekhov}}\ and\ \bibinfo {author} {\bibfnamefont {V.~P.}\ \bibnamefont
  {Gusynin}},\ }\bibfield  {title} {\bibinfo {title} {{RKKY} interaction in a
  doped pseudospin-1 fermion system at finite temperature},\ }\href
  {https://doi.org/10.1103/physrevb.101.235162} {\bibfield  {journal} {\bibinfo
   {journal} {Physical Review B}\ }\textbf {\bibinfo {volume} {101}},\ \bibinfo
  {pages} {235162} (\bibinfo {year} {2020})}\BibitemShut {NoStop}%
\bibitem [{\citenamefont {Yuan}\ \emph {et~al.}(2019)\citenamefont {Yuan},
  \citenamefont {Isobe},\ and\ \citenamefont {Fu}}]{Yuan2019}%
  \BibitemOpen
  \bibfield  {author} {\bibinfo {author} {\bibfnamefont {N.~F.~Q.}\
  \bibnamefont {Yuan}}, \bibinfo {author} {\bibfnamefont {H.}~\bibnamefont
  {Isobe}},\ and\ \bibinfo {author} {\bibfnamefont {L.}~\bibnamefont {Fu}},\
  }\bibfield  {title} {\bibinfo {title} {Magic of high-order van {H}ove
  singularity},\ }\href {https://doi.org/10.1038/s41467-019-13670-9} {\bibfield
   {journal} {\bibinfo  {journal} {Nature Communications}\ }\textbf {\bibinfo
  {volume} {10}},\ \bibinfo {pages} {5769} (\bibinfo {year}
  {2019})}\BibitemShut {NoStop}%
\bibitem [{\citenamefont {Shankar}\ \emph {et~al.}(2023)\citenamefont
  {Shankar}, \citenamefont {Oriekhov}, \citenamefont {Mitchell},\ and\
  \citenamefont {Fritz}}]{Shankar2023}%
  \BibitemOpen
  \bibfield  {author} {\bibinfo {author} {\bibfnamefont {A.~S.}\ \bibnamefont
  {Shankar}}, \bibinfo {author} {\bibfnamefont {D.~O.}\ \bibnamefont
  {Oriekhov}}, \bibinfo {author} {\bibfnamefont {A.~K.}\ \bibnamefont
  {Mitchell}},\ and\ \bibinfo {author} {\bibfnamefont {L.}~\bibnamefont
  {Fritz}},\ }\bibfield  {title} {\bibinfo {title} {Kondo effect in twisted
  bilayer graphene},\ }\href {https://doi.org/10.1103/physrevb.107.245102}
  {\bibfield  {journal} {\bibinfo  {journal} {Physical Review B}\ }\textbf
  {\bibinfo {volume} {107}},\ \bibinfo {pages} {245102} (\bibinfo {year}
  {2023})}\BibitemShut {NoStop}%
\bibitem [{\citenamefont {Saremi}(2007)}]{Saremi2007}%
  \BibitemOpen
  \bibfield  {author} {\bibinfo {author} {\bibfnamefont {S.}~\bibnamefont
  {Saremi}},\ }\bibfield  {title} {\bibinfo {title} {{RKKY} in half-filled
  bipartite lattices: Graphene as an example},\ }\href
  {https://doi.org/10.1103/physrevb.76.184430} {\bibfield  {journal} {\bibinfo
  {journal} {Physical Review B}\ }\textbf {\bibinfo {volume} {76}},\ \bibinfo
  {pages} {184430} (\bibinfo {year} {2007})}\BibitemShut {NoStop}%
\bibitem [{\citenamefont {Sherafati}\ and\ \citenamefont
  {Satpathy}(2011{\natexlab{a}})}]{Sherafati2011}%
  \BibitemOpen
  \bibfield  {author} {\bibinfo {author} {\bibfnamefont {M.}~\bibnamefont
  {Sherafati}}\ and\ \bibinfo {author} {\bibfnamefont {S.}~\bibnamefont
  {Satpathy}},\ }\bibfield  {title} {\bibinfo {title} {{RKKY} interaction in
  graphene from the lattice {G}reen’s function},\ }\href
  {https://doi.org/10.1103/physrevb.83.165425} {\bibfield  {journal} {\bibinfo
  {journal} {Physical Review B}\ }\textbf {\bibinfo {volume} {83}},\ \bibinfo
  {pages} {165425} (\bibinfo {year} {2011}{\natexlab{a}})}\BibitemShut
  {NoStop}%
\bibitem [{\citenamefont {Sherafati}\ and\ \citenamefont
  {Satpathy}(2011{\natexlab{b}})}]{Sherafati2011a}%
  \BibitemOpen
  \bibfield  {author} {\bibinfo {author} {\bibfnamefont {M.}~\bibnamefont
  {Sherafati}}\ and\ \bibinfo {author} {\bibfnamefont {S.}~\bibnamefont
  {Satpathy}},\ }\bibfield  {title} {\bibinfo {title} {Analytical expression
  for the {RKKY} interaction in doped graphene},\ }\href
  {https://doi.org/10.1103/physrevb.84.125416} {\bibfield  {journal} {\bibinfo
  {journal} {Physical Review B}\ }\textbf {\bibinfo {volume} {84}},\ \bibinfo
  {pages} {125416} (\bibinfo {year} {2011}{\natexlab{b}})}\BibitemShut
  {NoStop}%
\bibitem [{\citenamefont {Kogan}(2011)}]{Kogan2011}%
  \BibitemOpen
  \bibfield  {author} {\bibinfo {author} {\bibfnamefont {E.}~\bibnamefont
  {Kogan}},\ }\bibfield  {title} {\bibinfo {title} {{RKKY} interaction in
  graphene},\ }\href {https://doi.org/10.1103/physrevb.84.115119} {\bibfield
  {journal} {\bibinfo  {journal} {Physical Review B}\ }\textbf {\bibinfo
  {volume} {84}},\ \bibinfo {pages} {115119} (\bibinfo {year}
  {2011})}\BibitemShut {NoStop}%
\bibitem [{\citenamefont {Cao}\ \emph {et~al.}(2019)\citenamefont {Cao},
  \citenamefont {Fertig},\ and\ \citenamefont {Zhang}}]{CaoFertig2019}%
  \BibitemOpen
  \bibfield  {author} {\bibinfo {author} {\bibfnamefont {J.}~\bibnamefont
  {Cao}}, \bibinfo {author} {\bibfnamefont {H.~A.}\ \bibnamefont {Fertig}},\
  and\ \bibinfo {author} {\bibfnamefont {S.}~\bibnamefont {Zhang}},\ }\bibfield
   {title} {\bibinfo {title} {{RKKY} interactions in graphene {L}andau
  levels},\ }\href {https://doi.org/10.1103/PhysRevB.99.205430} {\bibfield
  {journal} {\bibinfo  {journal} {Phys. Rev. B}\ }\textbf {\bibinfo {volume}
  {99}},\ \bibinfo {pages} {205430} (\bibinfo {year} {2019})}\BibitemShut
  {NoStop}%
\bibitem [{\citenamefont {Fritz}\ and\ \citenamefont
  {Vojta}(2013)}]{Fritz2013}%
  \BibitemOpen
  \bibfield  {author} {\bibinfo {author} {\bibfnamefont {L.}~\bibnamefont
  {Fritz}}\ and\ \bibinfo {author} {\bibfnamefont {M.}~\bibnamefont {Vojta}},\
  }\bibfield  {title} {\bibinfo {title} {The physics of {K}ondo impurities in
  graphene},\ }\href {https://doi.org/10.1088/0034-4885/76/3/032501} {\bibfield
   {journal} {\bibinfo  {journal} {Reports on Progress in Physics}\ }\textbf
  {\bibinfo {volume} {76}},\ \bibinfo {pages} {032501} (\bibinfo {year}
  {2013})}\BibitemShut {NoStop}%
\bibitem [{{\relax DLMF}()}]{NIST:DLMF}%
  \BibitemOpen
  \bibfield  {author} {{\relax DLMF},\ }\bibfield  {title} {\bibinfo {title}
  {{\it NIST Digital Library of Mathematical Functions}},\ }\href
  {https://dlmf.nist.gov/} {\ }\bibinfo {note} {F.~W.~J. Olver, A.~B. {Olde
  Daalhuis}, D.~W. Lozier, B.~I. Schneider, R.~F. Boisvert, C.~W. Clark, B.~R.
  Miller, B.~V. Saunders, H.~S. Cohl, and M.~A. McClain, eds.,
  \url{https://dlmf.nist.gov/}}\BibitemShut {NoStop}%
\bibitem [{\citenamefont {Klier}\ \emph {et~al.}(2015)\citenamefont {Klier},
  \citenamefont {Shallcross}, \citenamefont {Sharma},\ and\ \citenamefont
  {Pankratov}}]{Klier2015}%
  \BibitemOpen
  \bibfield  {author} {\bibinfo {author} {\bibfnamefont {N.}~\bibnamefont
  {Klier}}, \bibinfo {author} {\bibfnamefont {S.}~\bibnamefont {Shallcross}},
  \bibinfo {author} {\bibfnamefont {S.}~\bibnamefont {Sharma}},\ and\ \bibinfo
  {author} {\bibfnamefont {O.}~\bibnamefont {Pankratov}},\ }\bibfield  {title}
  {\bibinfo {title} {{R}uderman-{K}ittel-{K}asuya-{Y}osida interaction at
  finite temperature: Graphene and bilayer graphene},\ }\href
  {https://doi.org/10.1103/PhysRevB.92.205414} {\bibfield  {journal} {\bibinfo
  {journal} {Phys. Rev. B}\ }\textbf {\bibinfo {volume} {92}},\ \bibinfo
  {pages} {205414} (\bibinfo {year} {2015})}\BibitemShut {NoStop}%
\bibitem [{\citenamefont {Vertogen}\ and\ \citenamefont
  {Caspers}(1966)}]{Vertogen1966}%
  \BibitemOpen
  \bibfield  {author} {\bibinfo {author} {\bibfnamefont {G.}~\bibnamefont
  {Vertogen}}\ and\ \bibinfo {author} {\bibfnamefont {W.~J.}\ \bibnamefont
  {Caspers}},\ }\bibfield  {title} {\bibinfo {title} {Contact part of the
  hyperfine interaction and the {R}uderman-{K}ittel-{K}asuya-{Y}osida
  approximation},\ }\href {https://doi.org/10.1103/physrevlett.16.904}
  {\bibfield  {journal} {\bibinfo  {journal} {Physical Review Letters}\
  }\textbf {\bibinfo {volume} {16}},\ \bibinfo {pages} {904} (\bibinfo {year}
  {1966})}\BibitemShut {NoStop}%
\bibitem [{\citenamefont {Bowen}(1968)}]{Bowen1968}%
  \BibitemOpen
  \bibfield  {author} {\bibinfo {author} {\bibfnamefont {S.~P.}\ \bibnamefont
  {Bowen}},\ }\bibfield  {title} {\bibinfo {title} {Divergences and phonons in
  the indirect interaction},\ }\href
  {https://doi.org/10.1103/physrevlett.20.726} {\bibfield  {journal} {\bibinfo
  {journal} {Physical Review Letters}\ }\textbf {\bibinfo {volume} {20}},\
  \bibinfo {pages} {726} (\bibinfo {year} {1968})}\BibitemShut {NoStop}%
\bibitem [{\citenamefont {Rusin}\ and\ \citenamefont
  {Zawadzki}(2020)}]{Rusin2020}%
  \BibitemOpen
  \bibfield  {author} {\bibinfo {author} {\bibfnamefont {T.~M.}\ \bibnamefont
  {Rusin}}\ and\ \bibinfo {author} {\bibfnamefont {W.}~\bibnamefont
  {Zawadzki}},\ }\bibfield  {title} {\bibinfo {title} {Exact {G}reen\'{}s
  function approach to {RKKY} interactions},\ }\href
  {https://doi.org/10.1103/physrevb.101.205201} {\bibfield  {journal} {\bibinfo
   {journal} {Physical Review B}\ }\textbf {\bibinfo {volume} {101}},\ \bibinfo
  {pages} {205201} (\bibinfo {year} {2020})}\BibitemShut {NoStop}%
\bibitem [{\citenamefont {Laubscher}\ \emph {et~al.}(2023)\citenamefont
  {Laubscher}, \citenamefont {Weber}, \citenamefont {H\"unenberger},
  \citenamefont {Schoeller}, \citenamefont {Kennes}, \citenamefont {Loss},\
  and\ \citenamefont {Klinovaja}}]{Klinovaja2023}%
  \BibitemOpen
  \bibfield  {author} {\bibinfo {author} {\bibfnamefont {K.}~\bibnamefont
  {Laubscher}}, \bibinfo {author} {\bibfnamefont {C.~S.}\ \bibnamefont
  {Weber}}, \bibinfo {author} {\bibfnamefont {M.}~\bibnamefont
  {H\"unenberger}}, \bibinfo {author} {\bibfnamefont {H.}~\bibnamefont
  {Schoeller}}, \bibinfo {author} {\bibfnamefont {D.~M.}\ \bibnamefont
  {Kennes}}, \bibinfo {author} {\bibfnamefont {D.}~\bibnamefont {Loss}},\ and\
  \bibinfo {author} {\bibfnamefont {J.}~\bibnamefont {Klinovaja}},\ }\bibfield
  {title} {\bibinfo {title} {{RKKY} interaction in one-dimensional flat-band
  lattices},\ }\href {https://doi.org/10.1103/PhysRevB.108.155429} {\bibfield
  {journal} {\bibinfo  {journal} {Phys. Rev. B}\ }\textbf {\bibinfo {volume}
  {108}},\ \bibinfo {pages} {155429} (\bibinfo {year} {2023})}\BibitemShut
  {NoStop}%
\bibitem [{sup()}]{supplement-code}%
  \BibitemOpen
  \href@noop {} {}\bibinfo {note} {The code for all numerical results presented
  in the paper can be found at the following zenodo
  \href{https://doi.org/10.5281/zenodo.15591563}{doi:10.5281/zenodo.15591563}
  repository}\BibitemShut {NoStop}%
\end{thebibliography}%
\end{document}